\documentclass[preprint,12pt]{elsarticle}
\usepackage{amsfonts}
\usepackage{amsmath}

\usepackage{amssymb}
\begin{document}

\begin{frontmatter}



\title{A semi-analytical finite element method for a class of
time-fractional diffusion equations}


\author{H. G.  Sun $^{a,b}$, W. Chen$^{b,*}$, K. Y. Sze$^{a}$}

\address{a.Department of Mechanical Engineering, The University of Hong Kong, Pokfulam, Hong
Kong SAR, P.R.China\\
 b.Institute of Soft Matter Mechanics, Department of
Engineering Mechanics, Hohai University, Nanjing, P.R.China\\
* Corresponding author: chenwen@hhu.edu.cn}

\begin{abstract}

As fractional diffusion equations can describe the early
breakthrough and the heavy-tail decay features observed in anomalous
transport of contaminants in groundwater and porous soil, they have
been commonly employed in the related mathematical descriptions.
These models usually involve long-time range computation, which is a
critical obstacle for its application, improvement of the
computational efficiency is of great significance. In this paper, a
semi-analytical method is presented for solving a class of
time-fractional diffusion equations which overcomes the critical
long-time range computation problem of time fractional differential
equations. In the procedure, the spatial domain is discretized by
the finite element method which reduces the fractional diffusion
equations into approximate fractional relaxation equations. As
analytical solutions exist for the latter equations, the burden
arising from long-time range computation can effectively be
minimized. To illustrate its efficiency and simplicity, four
examples are presented. In addition, the method is employed to solve
the time-fractional advection-diffusion equation characterizing the
bromide transport process in a fractured granite aquifer. The
prediction closely agrees with the experimental data and the
heavy-tail decay of anomalous transport process is well-represented.

\end{abstract}


\end{frontmatter}


\section{Introduction}

For the contaminant transport processes in soil and groundwater,
diffusion equations (such as diffusion equation,
advection-dispersion equation and advection-reaction-diffusion
equation) are the traditional governing equations
\cite{Dagan1987,LaBolle2001,Zhang2009,Metzler2000}. In the past
several decades, however, more and more evidences show that some of
the critical features in contaminant transport through complex
porous media cannot be described by the conventional diffusion
equations
\cite{Huang2008,Berkowitz1995,Zhang2007,Sun2009,Meerschaert2001,Benson2004}.
These features include early breakthrough and heavy-tail decay of
the contaminant as well as the scale-dependent coefficients
\cite{Berkowitz2006,Benson2000,Seymour2007}. They have led to the
increasing use of the fractional diffusion equations for modeling
contaminant transport in porous media.

The theoretical research on fractional diffusion equation models has
received considerable success in physical modeling and experimental
result analysis in the last decade
\cite{Metzler2000,Sokolov2005,Zaslavsky2002,Magin2008}.
Nevertheless, numerical methods for solving fractional diffusion
equations are still immature for practical applications in which the
spatial problem domains are geometrically complex and large whilst
the long-time range predictions are often desired. The main obstacle
is that the fractional derivative has its global nature, compared
with the locally expressed classic derivatives, the computational
cost of time fractional derivative term will increase dramatically
with time evolution. It brings computational challenge of
approximating fractional order equations with the finite difference
or the finite element methods \cite{Podlubny2009,Roop2006,Li2011}.
Though short memory method is proposed to tackle the long-time range
computation, it has been proved to bring accuracy deterioration in
many cases \cite{Podlubny1999,Ford2001}.
 Moreover, when computing high P$\acute{e}$clet number
problems, numerical schemes may produce oscillating solutions
\cite{Radu2011,Zienkiewicz2000,Lewis1996,Bergheau2008,Smith2004}.
The problem is more annoying in fractional advection-dispersion
equations.

Until now, the finite difference method has been widely employed for
solving anomalous diffusion equations with successes in short-time
and small spatial scale problems \cite{Yuste2006,Chen2007}. Since
the finite element method (FEM) is more suitable for modeling large
and geometrically complicated spatial domains, the investigation of
the FEM for fractional diffusion type equations has attracted much
attention in recent years. Roop et al. approximated the solutions of
steady state space-fractional advection-dispersion equations in two
spatial dimensions using the Galerkin and least-squares FEMs
\cite{Roop2006, Fix2004}. Huang et al. proposed an unconditionally
stable FEM approach to solve the one-dimensional space-fractional
advection-dispersion equation, and successfully applied it to
simulate the atrazine transport in a saturated soil column
\cite{Huang2008}. Deng developed a FEM for the numerical resolution
of the space and time-fractional Fokker-Planck equation with its
convergence order is $O(k^{2-\alpha}+h^\mu)$, $\alpha$ and $\mu$ are
time and spatial derivative orders \cite{Deng2008}. Zhuang et al.
investigated the Galerkin finite element approximation of symmetric
space-fractional partial differential equations, and proved the
stability and convergence of the proposed schemes \cite{Zhang2010}.
Zheng et al. discussed the FEM for the space-fractional advection
diffusion equation with non-homogeneous initial-boundary condition
\cite{Zheng2010,Zheng2010b,Li2011}. Though all aforementioned works
indicate that FEMs play an important and increasing role in the
applications of fractional diffusion type equation models, the
efficient and simple numerical methods for fractional diffusion
equations are still urgently needed.

In this paper, we introduce a semi-analytical FEM for
time-fractional diffusion equations which can be expressed in the
following form:
\begin{eqnarray}
\begin{array}{c}
\displaystyle{\frac{d^\gamma u}{d t^\gamma} =-\textbf{A}^T  \nabla u+ \nabla^T  (\textbf{D} \nabla u)+ P u+ f,\,\,0<\gamma \leq 1} \\
\end{array}
\label{eq0}
\end{eqnarray}
in which $u$ is the scalar unknown, $t$ denotes time, $\gamma$ is
the fractional derivative order, $\nabla$ is the gradient operator,
$A$ is a coefficient vector, $D$ is a coefficient matrix, $P$ and
$f$ are scalars. Moreover, $A$, $D$, $P$  and $f$ are functions of
the spatial variables. In the proposed semi-analytical FEM, the FEM
is employed for spatial discretization which reduces the
time-fractional diffusion equations to approximate fractional
relaxation equations (also known as the temporal fractional ODEs).
The FEM can conveniently be used to discretize large and
geometrically complicated spatial domains. On the other hand, the
analytical solutions for fractional relaxation equations exist and
the burden of long-time range computation can be significantly
alleviated.  The present semi-analytical method can solve 1D, 2D and
3D problems with variable coefficients conveniently at low
implementation cost.

%in which $u$ is contaminant concentration, $\nabla$ is the gradient
%operator, $\textbf{A}$ is the generalized convective coefficient
%vector, $\textbf{D}$ denotes the generalized diffusion coefficient
%matrix,  $P u$ represents a reaction-absorption term and $f$ is a
%source term. $A,\,D,\,P\,$ and $f$ are functions of the spatial
%variables. The fractional order $\frac{d^\gamma}{d t^\gamma} $ is
%designed to characterize the anomalous feature of the considered
%contaminant transport problem, the fractional order $\gamma$ is
%determined by the physical properties of porous medium. In the
%proposed semi-analytical FEM, the FEM is employed for spatial
%discretization which reduces the time-fractional diffusion equations
%to approximate fractional relaxation equations (also known as the
%temporal fractional ODEs). The FEM can conveniently be used to
%discretize large and geometrically complicated spatial domain. On
%the other hand, the analytical solutions for fractional relaxation
%equations exist and the burden of long-time range computation can be
%significantly eliminated. The present semi-analytical method can
%solve 1D, 2D and 3D problems with variable coefficients conveniently
%at low implementation cost.

%To prove the efficiency and accuracy of proposed method, we solve
%four academic examples using the proposed method, and further employ
%it to compute the time fractional transport equation for bromide
%transport in fractured granite aquifer.

\section{Algorithm framework}

Time-fractional diffusion equations are often used to characterize
the contaminant transport processes, which exhibit typical
subdiffusion features, such as heavy-tail decay of concentration and
nonlinear time dependent mean square displacement
 $\langle x^2(t)\rangle \sim t^\alpha\,(0<\alpha<1) $ \cite{Metzler2000}. The time-fractional diffusion problem can be expressed
 as:
\begin{eqnarray}
\left\{
\begin{array}{c}
\displaystyle{\frac{d^\gamma u}{d t^\gamma}=-\textbf{A}^T  \nabla u+ \nabla^T (\textbf{D} \nabla u)+ P u +f \,\,\,\,\text{in} \,\,\Omega \times (0, T),} \\
\displaystyle{u=\bar{u}\,\,\text{on}\,\, \Gamma_D,}\\
\displaystyle{\textbf{n}^T (\textbf{D}  \nabla u)=q\,\,\text{on}\,\, \Gamma_N,}\\
\displaystyle{\textbf{n}^T (\textbf{D}  \nabla u)=h_c(u-u_\infty)\,\,\text{on}\,\, \Gamma_C,}\\
\displaystyle{u|_{t=0}=u_0\,\,\text{in} \,\, \Omega}.
\end{array}\right.\;
\label{eq1}
\end{eqnarray}

Here, $u$ is contaminant concentration, $A$ is the generalized
convective coefficient vector, $D$ is the generalized diffusion
coefficient matrix, $Pu$ represents a reaction-absorption term, $f$
is the source term and $\Omega$ denotes the spatial domain of the
problem. Moreover, $\textbf{n}$ is the unit outward normal vector to
the boundary, $h_c$ is the convective coefficient and $\Gamma_D \cup
\Gamma_N \cup \Gamma_C=\partial \Omega$ which is denotes the entire
boundary of $\Omega$. The subscripts $D$, $N$ and $C$ for $\Gamma$
designate the essential (Dirichlet), natural (Neumann) and
convective boundary conditions, respectively, whereas $\Gamma_D$,
$\Gamma_N$ and $\Gamma_C$ are mutually exclusive. It will be assumed
that $A$, $D$, $P$, $f$, $q$ and $u_\infty$ are independent of time.
In the problem statement, $\frac{d^\gamma}{d t^\gamma}$ represents
the Caputo fractional derivative whose definition is given as below:
\begin{eqnarray}
\begin{array}{c}
\displaystyle{\frac{d^\gamma }{d t^\gamma} g(t)
=\frac{1}{\Gamma(1-\gamma)}\int_{0}^t\frac{g'(\tau){\rm
d}\tau}{(t-\tau)^{\gamma}} ,\, 0<\gamma \leq 1} \\
\end{array}
\label{eqc}
\end{eqnarray}
in which $\Gamma$ is the Gamma function and $\gamma$ is the
derivative order. The Caputo fractional derivative has the following
properties \cite{Samko1993,Kumar2006}:

\begin{eqnarray}
\left\{
\begin{array}{l}
\displaystyle{\frac{d^\gamma }{d t^\gamma} E_\gamma (\lambda t^\gamma)= \lambda E_\gamma (\lambda t^\gamma),} \\
\displaystyle{\frac{d^\gamma }{d t^\gamma} Constant = 0}
\end{array}\right.\;
\label{eqadd}
\end{eqnarray}
in which  $E_\gamma$ represents the Mittag-Leffler function with one
parameter \cite{Podlubny1999}:
\begin{eqnarray}
\begin{array}{c}
\displaystyle{E_\gamma (z)=\sum_{n=0}^\infty \frac{z^n}{\Gamma
(\gamma n+1)},\,\, \gamma>0,\,\, z\in \emph{\textbf{C}}.}
\end{array}
\label{eqml}
\end{eqnarray}
The weighted residual statements for the governing equation, natural
boundary condition and convective boundary condition:
\begin{eqnarray}
\left\{
\begin{array}{l}
\displaystyle{\int_\Omega \psi[ \frac{d^\gamma u}{d t^\gamma}+\textbf{A}^T  \nabla u- \nabla^T (\textbf{D} \nabla u)- P u -f] d \Omega=0,} \\
\displaystyle{\int_{\Gamma_N} \psi [\textbf{n}^T (\textbf{D} \nabla
u)-q] d \Gamma= 0,}\\
\displaystyle{\int_{\Gamma_C} \psi [\textbf{n}^T (\textbf{D} \nabla
u)-h_c (u-u_\infty)] d \Gamma= 0}
\end{array}\right.\;
\label{eqadd}
\end{eqnarray}
can be merged to form the following weak form with the help of the
divergence theorem:
\begin{eqnarray}
\begin{array}{c}
\displaystyle{\int_\Omega \psi \frac{d^\gamma u}{d t^\gamma} d \Omega+ \int_\Omega \psi \textbf{A}^T  (\nabla u) d \Omega+ \int_\Omega (\nabla \psi)  \textbf{D} (\nabla u) d \Omega- \int_\Omega P \psi   u d \Omega} \\
\displaystyle{=\int_{\Gamma_N}\psi q d \Gamma +\int_{\Gamma_C}\psi
h_c (u-u_\infty) d \Gamma+\int_\Omega \psi f d \Omega}
\end{array}
\label{eq2}
\end{eqnarray}
in which the trial solution $u$ equals $\bar{u}$ and the weight
function $\psi$ vanishes on $\Gamma_D$. In the finite element
method, $u$ and $\Psi$ within each element $\Omega^e$ can be
expressed respectively as:
\begin{eqnarray}
\left\{
\begin{array}{c}
\displaystyle{ u^e= \sum_{i=1}^{n-n_D} N_i^e u_i^e+\sum_{i=1}^{n_D} \bar{N}_i^e \bar{u}_i^e  =  \textbf{N}^e  \textbf{U}^e+\bar{\textbf{N}}^e  \bar{\textbf{U}}^e }, \\
\displaystyle{\psi^e= \sum_{i=1}^{n-n_D} N_i^e \psi_i^e =
\textbf{N}^e \Psi^e=(\Psi^e)^T (\textbf{N}^e)^T }
\end{array}\right.\;
\label{eq3}
\end{eqnarray}
where $u_i^e$ and $\bar{u}_i^e$ are the values of $u$ at nodes away
from and on $\Gamma_D$, respectively; $\psi_i^e$ are the value of
$\psi$ at nodes away from $\Gamma_D$; $N_i^e$ and $\bar{N}_i^e$ are
the nodal interpolation functions for $u_i^e$ and $\bar{u}_i^e$,
respectively; $n$ is the number of nodes in the element, $n_D$ is
the number of nodes on $\Gamma_D^e=\partial\Omega^e \cap \Gamma_D$,
$N_i^e$s and $\bar{N}_i^e$s form the row interpolation matrices
$\textbf{N}^e$ and $\bar{\textbf{N}}^e$, respectively; $u_i^e$s and
$\bar{u}_i^e$s form the vectors $\textbf{U}^e$ and
$\bar{\textbf{U}}^e$, respectively. By virtue of (\ref{eq3}),
(\ref{eq2}) can be written as:
\begin{eqnarray}
\displaystyle{\sum_e (\Psi^e)^T ([\textbf{C}^e, \bar{\textbf{C}}^e]
\frac{d^\gamma}{d t^\gamma}\left\{
\begin{array}{c}
\textbf{U}^e\\
\bar{\textbf{U}}^e
\end{array}
\right\} +[\textbf{K}^e, \bar{\textbf{K}}^e]} \left\{
\begin{array}{c}
\textbf{U}^e\\
\bar{\textbf{U}}^e
\end{array}
\right\}-\textbf{F}^e)=0, \label{eqS1}
\end{eqnarray}
or
\begin{eqnarray}
\displaystyle{(\Psi)^T ([\textbf{C}, \bar{\textbf{C}}]
\frac{d^\gamma}{d t^\gamma} \left\{
\begin{array}{c}
\textbf{U}\\
\bar{\textbf{U}}
\end{array}\right\} +[\textbf{K},
\bar{\textbf{K}}]} \left\{
\begin{array}{c}
\textbf{U}\\
\bar{\textbf{U}}
\end{array}
\right\}-\textbf{F})=0 \label{eqS2}
\end{eqnarray}
in which
\begin{eqnarray}
\begin{array}{l}
\displaystyle{[\textbf{C}^e, \bar{\textbf{C}}^e]=\int_{\Omega^e}
(\textbf{N}^e)^T [\textbf{N}^e, \bar{\textbf{N}}^e] d \Omega},\\
\displaystyle{[\textbf{K}^e, \bar{\textbf{K}}^e]=\int_{\Omega^e}
((\textbf{N}^e)^T \textbf{A}^T [\nabla \textbf{N}^e, \nabla
\bar{\textbf{N}}^e] +(\nabla \textbf{N}^e)^T \textbf{D} [\nabla
\textbf{N}^e, \nabla \bar{\textbf{N}}^e]}\\
\displaystyle{\,\,\,\,\,\,\,\,\,\,\,\,\,\,\,\,\,\,\,\,\,\,\,\,\,\,\,\,-
P (\textbf{N}^e)^T [\nabla \textbf{N}^e, \nabla \bar{\textbf{N}}^e])
d \Omega-\int_{\Gamma_C^e} h_c (\textbf{N}^e)^T [\textbf{N}^e,
\bar{\textbf{N}}^e] d \Gamma},\\
\displaystyle{\textbf{F}^e=\int_{\Gamma_N^e} q (\textbf{N}^e)^T
 d \Gamma-\int_{\Gamma_C^e} h_c u_{\infty} (\textbf{N}^e)^T d \Gamma+\int_{\Omega^e}
f (\textbf{N}^e)^T  d \Omega}.
\end{array}
\label{eqS3}
\end{eqnarray}
In $F^e$, $\Gamma_N^e=\partial \Omega^e \cap \Gamma_N$ and
$\Gamma_C^e=\partial \Omega^e \cap \Gamma_C$. Moreover, $\Psi,$
$\textbf{C},$ $\bar{\textbf{C}},$ $\textbf{K},$ $ \bar{\textbf{K}},$
$ \textbf{U},$ $ \bar{\textbf{U}}$ and $\textbf{F}$ are the
assembled counterparts of $\Psi^e,$ $\textbf{C}^e,$
$\bar{\textbf{C}}^e,$ $ \textbf{K}^e,$ $ \bar{\textbf{K}}^e,$ $
\textbf{U}^e,$ $ \bar{\textbf{U}}^e$ and $\textbf{F}^e$,
respectively.  The arbitrary nature of $\Psi$ leads to the following
system equation:
\begin{eqnarray}
\displaystyle{\textbf{C}\frac{d^\gamma}{d
t^\gamma}\textbf{U}+\textbf{K}
\textbf{U}+\bar{\textbf{C}}\frac{d^\gamma}{d
t^\gamma}\bar{\textbf{U}}+\bar{\textbf{K}}
\bar{\textbf{U}}-\textbf{F}=0}. \label{eqS4}
\end{eqnarray}
%Noting that $\bar{\textbf{K}} \bar{\textbf{U}}-\textbf{F}$ is
%independent of time, the above system equation can be transformed
%into:
If
\begin{eqnarray}
\displaystyle{\bar{\textbf{C}}\frac{d^\gamma}{d
t^\gamma}\bar{\textbf{U}}+\bar{\textbf{K}}
\bar{\textbf{U}}=\bar{\textbf{F}}\neq \bar{\textbf{F}}(t)},
\label{eqS7}
\end{eqnarray}
the above system equation can be transformed into:
\begin{eqnarray}
\displaystyle{\textbf{C}\frac{d^\gamma}{d
t^\gamma}\tilde{\textbf{U}}+\textbf{K} \tilde{\textbf{U}}=0}
\label{eqS5}
\end{eqnarray}
where
\begin{eqnarray}
\displaystyle{\tilde{\textbf{U}}=\textbf{U}+\textbf{K}^{-1}(\bar{\textbf{K}}
\bar{\textbf{U}}+\bar{\textbf{F}}-\textbf{F})}. \label{eqS6}
\end{eqnarray}
At last,  the time-fractional system (\ref{eq1}) leads to:
\begin{eqnarray}\left\{
\begin{array}{c}
\displaystyle{\textbf{C}\frac{d^\gamma }{d t^\gamma}
\tilde{\textbf{U}}+\textbf{K}
\tilde{\textbf{U}}= 0},\\
\displaystyle{\tilde{\textbf{U}}|_{t=0}= \tilde{\textbf{U}}_0}.
\end{array}\right.\;
\label{eq8}
\end{eqnarray}
Exact solution of the fractional relaxation equation in (\ref{eq8})
exists and can be expressed as \cite{Guymon1970,Mainardi1996}:
\begin{eqnarray}
\begin{array}{c}
\displaystyle{\tilde{\textbf{U}}_t= E_\gamma (-\textbf{M}t^\gamma)}
\tilde{\textbf{U}}_0
\end{array}
\label{eq9}
\end{eqnarray}
where $\textbf{M}=\textbf{C}^{-1}\textbf{K}$, $E_\gamma$ is the
Mittag-Leffler function which has been accurately evaluated by
Podlubny et al \cite{Podlubny2009a,Chen2008}. In our computations,
the employed value of the function will be accurate up to
$10^{-12}$.

It can be deduced that, if $\gamma=1.0$, (\ref{eq9}) becomes the
exponential solution of the integer order relaxation equation. Next,
we decompose the Mittag-Leffler function in (\ref{eq9})  as:
\begin{eqnarray}
\begin{array}{c}
\displaystyle{E_\gamma (-\textbf{M} t^\gamma)=\textbf{B} \Lambda_t
\textbf{B}^{-1}}
\end{array}
\label{eq10}
\end{eqnarray}
where $\textbf{B}$ is the modal matrix formed by the eigenvectors of
$-\textbf{M}$. On the other hand, $\Lambda_t$ is a diagonal matrix
whose $i$-th diagonal entries is $E_\gamma (-\Lambda_i t^\gamma)$
and $\Lambda_i$ is the $i$-th eigenvalue of $-\textbf{M}$.
Substituting (\ref{eq10}) into (\ref{eq9}), we get
\begin{eqnarray}
\begin{array}{c}
\displaystyle{\tilde{\textbf{U}}_t= \textbf{B} \Lambda_t
\textbf{B}^{-1} \tilde{\textbf{U}}_0}.
\end{array}
\label{eq11}
\end{eqnarray}

Through the above manipulations, the initial-boundary value problem
in (\ref{eq1}) is reduced to a initial problem through spatial
finite element discretization. The reduced problem can be solved
analytically in terms of the Mittag-Leffler function. The practice
drastically lowers the cost associated with the long-time range
computation of the initial-boundary value problem.

%transformations, the solution of (\ref{eq1}) can be approximated by
%computing (\ref{eq11}). It means that the FEM is adopted in spatial
%discretization, and the exact solution is performed in time
%direction. In addition, the formulas (\ref{eq10})-(\ref{eq11}) have
%dramatically decreased the computational cost of matrix, it is
%really important in the computation of two or three dimensional
%problems. It can be seen from (\ref{eq11}) that when computing a
%numerical value $u(*,t)$, we only need to know the initial value
%$u(*,0)$, hence the new method has a rather low computational cost
%in time direction. We should also point out that, if the essential
%boundary condition is $u|_{\Gamma_D}=E_\gamma (-m t^\gamma) u_0$ (
%where $E_\gamma (-m t^\gamma)$ is the typical decay function) which
%is consistent with the time decay feature of the system, it can also
%be tackled by the proposed method, we will further illustrate it by
%numerical experiments.
For the essential boundary condition in (\ref{eqS7}), the equation
can be transformed as:
\begin{eqnarray}
\displaystyle{\bar{\textbf{C}}\frac{d^\gamma}{d
t^\gamma}(\bar{\textbf{U}}-\bar{\textbf{K}}^{-1}\bar{\textbf{F}})+\bar{\textbf{K}}
(\bar{\textbf{U}}-\bar{\textbf{K}}^{-1}\bar{\textbf{F}})=0}
\label{eqS8}
\end{eqnarray}
which would require
$\bar{\textbf{U}}-\bar{\textbf{K}}^{-1}\bar{\textbf{F}}$ to satisfy
a fractional relaxation equation. In particular, the typical case in
which $\bar{\textbf{U}}$ is a constant $d^\gamma \bar{\textbf{U}}/d
t^\gamma=0$ also satisfies (\ref{eqS7}).

\section{Numerical examples}
\subsection{One dimensional time-fractional diffusion equation}
\begin{figure}
\begin{center}
\includegraphics[width=0.8\textwidth]{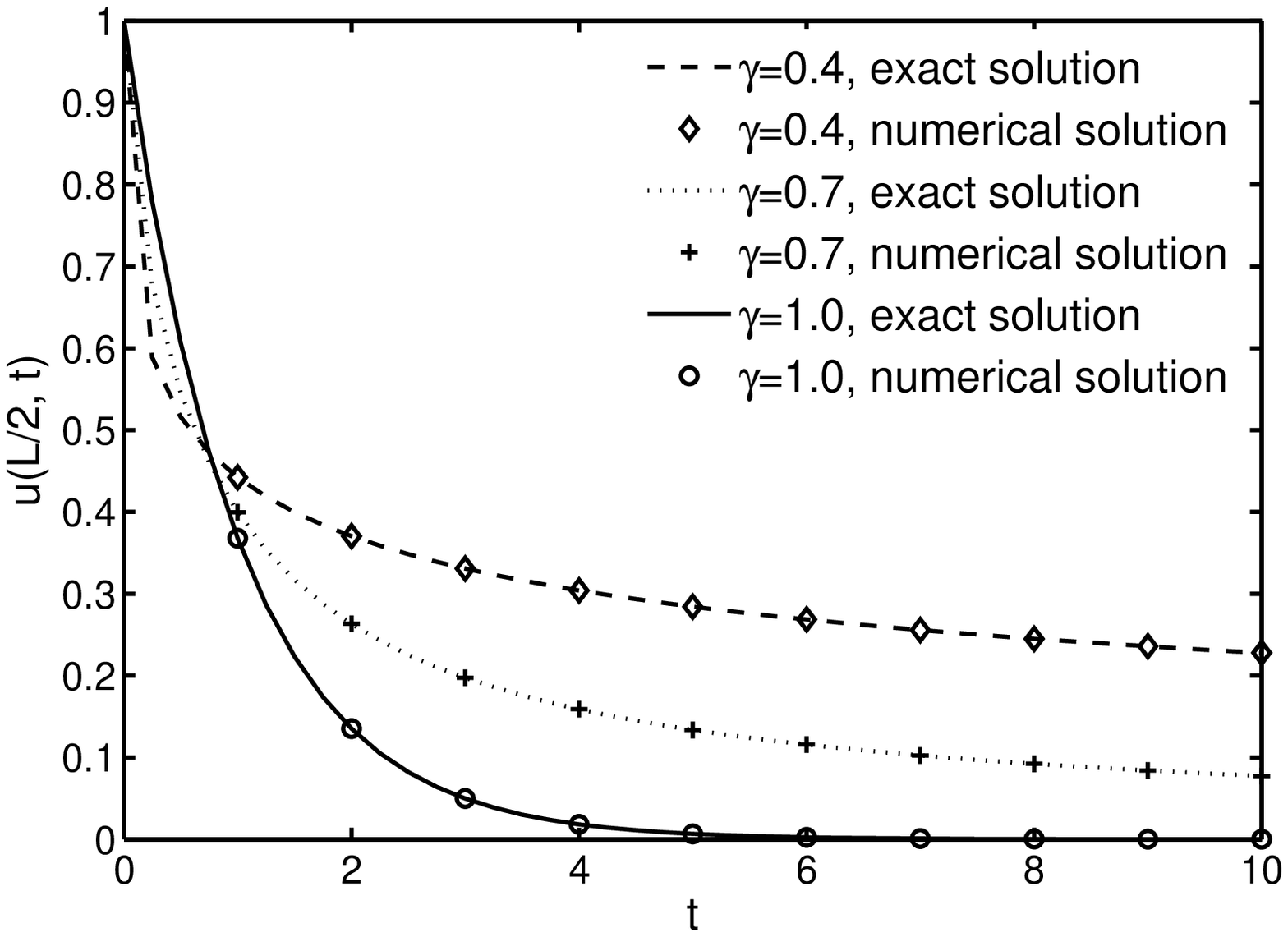}% Here is how to import EPS art
\end{center}\caption{\label{fig:epsart} The diffusion curves of time-fractional
diffusion model (\ref{eq12}) at $x=L/2$, obtained by the quadratic
element. Nodal spacing $h=L/100$ and space size $L=10$. }
\label{fig1}
\end{figure}

The following simple time-fractional diffusion problem is
considered:
\begin{eqnarray}
\left\{
\begin{array}{c}
\displaystyle{\frac{d^\gamma u(x, t)}{d t^\gamma}= k \frac{\partial^2 u(x,t)}{\partial x^2},\,\,x\in (0,L),\,\,t>0,} \\
\displaystyle{u(0,t)=u(L,t)=0,}\\
u(x,0)=sin(\pi x/L),\,\,x\in [0,L].
\end{array}\right.\;
\label{eq12}
\end{eqnarray}
If the diffusion coefficient $k=L^2/ \pi^2$, the exact solution is
 $u_{exact}(x,t)=sin(\pi x/L) E_\gamma (-t^\gamma)$ whose value at $x=L/2$ is shown in Figure \ref{fig1} for $\gamma=0.4, 0.7$ and $1.0$.

To construct the spatial discretization, both linear and quadratic
elements are attempted. For the linear element, the shape functions
are:
\begin{eqnarray}
\begin{array}{c}
\displaystyle{N_1(\xi)=(1-\xi)/2,\,\,N_2(\xi)=(1+\xi)/2,\,\, \xi \in
[-1,\,1]},
\end{array}
\label{eq13}
\end{eqnarray}
and the element matrices are:
\begin{eqnarray}
\textbf{C}^e=\frac{h}{6}\left[
\begin{array}{cc}
     2 & 1\\
     1 & 2
\end{array}\right],\;
\textbf{K}^e=\frac{k}{h}\left[
\begin{array}{cc}
     1 & -1\\
     -1 & 1
\end{array}\right]\;
\label{eq14}
\end{eqnarray}
in which $h$ is the nodal spacing. For the quadratic element, the
shape functions are:
\begin{eqnarray}
\begin{array}{c}
\displaystyle{N_1(\xi)=-\xi(1-\xi)/2,\,\,N_2(\xi)=(1-\xi^2),\,\,N_3(\xi)=\xi(1+\xi)/2},
\end{array}
\label{eq15}
\end{eqnarray}
and the corresponding element matrices are:
\begin{eqnarray}
\textbf{C}^e=\frac{h}{30}\left[
\begin{array}{ccc}
     4 & 2 & -1\\
     2 & 16 & 2\\
     -1 & 2 & 4
\end{array}\right],\;
\textbf{K}^e=\frac{k}{3 h}\left[
\begin{array}{ccc}
     7 & -8 & 1\\
     -8 & 16 & -8\\
     1 & -8 & 7
\end{array}\right].\;
\label{eq16}
\end{eqnarray}

Table \ref{tab1} lists the normalized errors at different time
instants yielded by using $10$ linear, $10$ quadratic and $100$
linear elements. The proposed method can achieve accurate results no
matter linear or or quadratic elements are employed. As usual, the
quadratic element delivers much more accurate results than the
linear element at the same nodal spacing. Another important feature
of this method is that the accuracy of numerical result at large
time constants can be improved by reducing the nodal spacing $h$.

%With the above element matrices, we can obtain numerical result of
%(\ref{eq12}) by following the calculation steps
%(\ref{eq8})-(\ref{eq11}). The corresponding numerical result is
%shown in Figure \ref{fig1}, the normalized error result is stated in
%Table \ref{tab1}. From the observation of Table \ref{tab1}, we can
%claim that the proposed method can achieve accurate numerical result
%no matter by the linear element or the quadratic element. In
%addition, the quadratic element gives more accurate numerical result
%than the linear element. Another important feature of this method is
%that the accuracy of numerical result at large time constants can be
%improved by decreasing the nodal spacing $h$.

  \begin{table}
  \centering
   \caption{A comparison of normalized errors (Error$=|(u_{exact}(L/2, t)-u(L/2, t))/u_{exact}(L/2, t)|$) of the linear element and the quadratic element.
Space size $L=10$, time-fractional derivative order $\gamma=0.8$ and
diffusion coefficient $k=L^2/\pi^2$ in (\ref{eq12}).}
\begin{tabular}{c  c  c  c}
   \hline
   Time & Linear element & Quadratic element
 &  Linear element  \\
     & ($h=L/10$) &
($h=L/10$) &   ($h=L/100$) \\
   \hline
  t=0.0 & 0.00000000 & 0.00000000 & 0.00000000 \\
  t=0.1 & 1.3517e-003 & 1.0525e-006 & 1.3485e-005 \\
  t=0.2 & 2.2855e-003 & 0.4709e-006 & 2.2814e-005 \\
  t=0.3 & 3.0766e-003 & 1.7646e-006 & 3.0726e-005 \\
  t=0.4 & 3.7734e-003 & 2.9057e-006 & 3.7703e-005 \\
  t=0.5 & 4.3983e-003 & 3.9302e-006 & 4.3965e-005 \\
  t=0.6 & 4.9644e-003 & 4.8592e-006 & 4.9643e-005 \\
  t=0.7 & 5.4805e-003 & 5.7070e-006 & 5.4825e-005 \\
  t=0.8 & 5.9530e-003 & 6.4838e-006 & 5.9572e-005 \\
  t=0.9 & 6.3868e-003 & 7.1975e-006 & 6.3934e-005 \\
   \hline
 \end{tabular}
 \label{tab1}
  \end{table}

To estimate the convergence ratio of the linear element and the
quadratic element, the results in Table \ref{tab3} evaluated at
$t=10$ but different nodal spacings are prepared and the
$L_\infty$-error is:
\begin{eqnarray}
\begin{array}{c}
\displaystyle{L_{\infty, h}=\max_i
|u_{exact}(x(i),t)-u(x(i),t)|,\,\,i=1,2,...,L/h,}
\end{array}
\label{eqmax}
\end{eqnarray}
It can be seen that the convergence ratio of the linear element is
$O(h^2)$ and the quadratic element is $O(h^4)$.

In Table \ref{tab1}, the normalized errors increase with  $t$. To
investigate the efficiency in tackling long-time range diffusion
problems, the normalized errors of the linear and the quadratic
elements at large $t$ are computed and listed in Table \ref{tab10}.
It can be seen that the normalized errors remain fairly steady with
respect to $t$.
  \begin{table}
  \centering
   \caption{The $L_\infty$-errors and convergence ratios of the linear element and the quadratic element ($Ratio=log(L_{\infty, h_1}/L_{\infty, h_2}) [log(h_1/h_2)]^{-1}$) \cite{Su2009}.
   Time-fractional derivative order
$\gamma=0.8$, space size $L=10$, diffusion coefficient
 $k=L^2/\pi^2$ and $t=10$ in (\ref{eq12}).}
\begin{tabular}{c c c c c}
   \hline
   Nodal spacing & $L_\infty$-error & Ratio
 &  $L_\infty$-error & Ratio \\
  & (Linear element) &
 &  (Quadratic element) &   \\
   \hline
  h=L/10 & 4.327591e-004 &            & 7.739342e-007 & \\
  h=L/20 & 1.087320e-004 & 1.9928 & 4.909022e-008 & 3.9787\\
  h=L/40 & 2.721688e-005 & 1.9982 & 3.080541e-009 & 3.9942\\
  h=L/80 & 6.806336e-006 & 1.9996 & 1.937557e-010 & 3.9909\\
  h=L/160 & 1.701717e-006 & 1.9999 & 1.265827e-011 & 3.9361\\
   \hline
 \end{tabular}
 \label{tab3}
  \end{table}

  \begin{table}
  \centering
   \caption{The normalized errors (Error$=|(u_{exact}(L/2, t)-u(L/2, t))/u_{exact}(L/2, t)|$) of the quadratic element. Space size $L=10$, diffusion coefficient $k=L^2/\pi^2$, node
spacing $h=L/100$ and time-fractional derivative order $\gamma=0.8$
in (\ref{eq12}). }
\begin{tabular}{c  c  c  c c}
   \hline
   Methods & t=10 & t=100 &  t=1000 &  t=10000 \\
   \hline
  Linear element & 1.0136e-004 & 8.4909e-005 & 8.2652e-005 & 8.2307e-005 \\
  Quadratic element & 1.3185e-009 & 1.0614e-009 & 1.0242e-009 & 1.0186e-009 \\
   \hline
 \end{tabular}
 \label{tab10}
  \end{table}

\subsection{One dimensional time-fractional convection-dispersion equation}
\begin{figure}
\begin{center}
\includegraphics[width=0.8\textwidth]{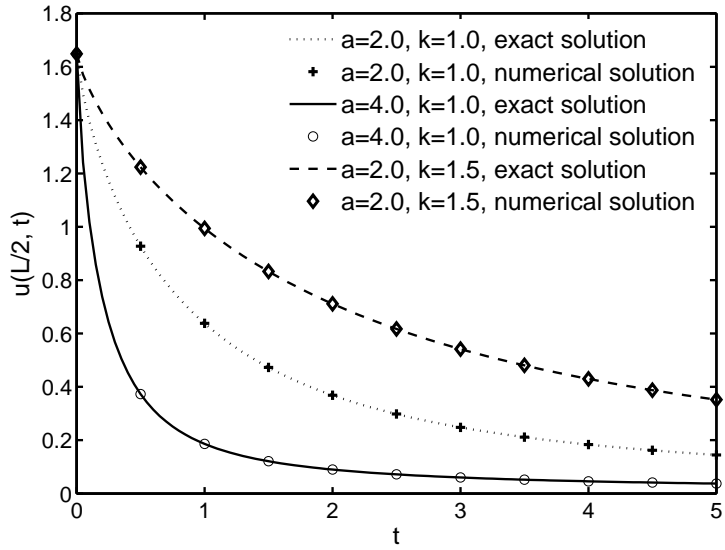}% Here is how to import EPS art
\end{center} \caption{\label{fig:epsart} A comparison of exact and numerical solutions of time-fractional advection-dispersion model (\ref{eq17})
at $x=L/2$ with $\gamma=0.8$. Nodal spacing $h=L/20$ and space size
$L=1.0$ in the linear element. } \label{fig2}
\end{figure}

Time-fractional advection-dispersion equation (also called
time-fractional Fokker-Planck equation), which exhibits heavy-tail
concentration decay feature, is usually used to characterize
contaminant transport in natural porous media. A simple example is:

\begin{eqnarray} \left\{
\begin{array}{c}
\displaystyle{\frac{d^\gamma u(x,t)}{d t^\gamma}=-a \frac{\partial u(x,t)}{\partial x}+ k \frac{\partial^2 u(x,t)}{\partial x^2},\,\,x\in (0,L),\,\,t>0,} \\
\displaystyle{u(0,t)=E_\gamma (-(a-k)t^\gamma),\,\, \,\, u(L,t)=e^L E_\gamma (-(a-k)t^\gamma),\,\,}\\
u(x,0)=e^x,\,\, x\in [0, L].
\end{array}\right.\;
\label{eq17}
\end{eqnarray}
Assuming $a$ and $k$ are constants, $a>k$, the exact solution of
above equation can be written as:
\begin{eqnarray}
\displaystyle{u_{exact}(x,t)=e^x E_\gamma (-(a-k)t^\gamma)}
\label{eq18}
\end{eqnarray}
which is portrayed in Figure \ref{fig2} for $t<1$. For the linear
element, the corresponding element matrices are:
\begin{eqnarray}
\textbf{C}^e=\frac{h}{6}\left[
\begin{array}{cc}
     2 & 1\\
     1 & 2
\end{array}\right],\;
\textbf{K}^e=\frac{a}{2}\left[
\begin{array}{cc}
     -1 & 1\\
     -1 & 1
\end{array}\right]\;
+\frac{k}{h}\left[
\begin{array}{cc}
     1 & -1\\
     -1 & 1
\end{array}\right].\;
\label{eq19}
\end{eqnarray}

The normalized errors obtained by using $10$, $20$ and $40$ elements
at $x = L/2$ and different $t$ are shown in Table \ref{tab2}. With
only 10 elements, the errors have been less than $0.1\%$.

\subsection{Two dimensional time-fractional diffusion equation}
In this example, the following two-dimensional problem is
considered:
\begin{eqnarray}
\left\{
\begin{array}{c}
\displaystyle{\frac{d^\gamma u(x,y,t)}{d t^\gamma} = k (\frac{\partial^2 u(x,y,t)}{\partial x^2}+ \frac{\partial^2 u(x,y,t)}{\partial y^2}),\,\,(x,y)\in \Omega,} \\
\displaystyle{u(x,y,t)=0,\,\,(x,y)\in \partial \Omega,\,\,t>0,}\\
u(x,y,0)=sin(x\pi/L)sin(y\pi/L),\,\,(x,y)\in \Omega \cup
\partial \Omega,
\end{array}\right.\;
\label{eq20}
\end{eqnarray}
in which $k$ is the diffusion coefficient, $\Omega=[0,L]\times
[0,L]$. If $k=1/\pi^2$ and $L=1.0$, the exact solution of the
problem is $u_{exact}(x,y,t)=sin(x\pi)sin(y\pi) E_\gamma (-2
t^\gamma)$.

  \begin{table}
  \centering
   \caption{The normalized errors (Error$=|(u_{exact}(L/2, t)-u(L/2, t))/u_{exact}(L/2, t)|$) of the linear element. Space size $L=1.0$, diffusion coefficient $k=1.0$,\,\, convective coefficient $a=2.0$ and time-fractional derivative order
$\gamma=0.8$ in (\ref{eq17}).}
\begin{tabular}{c c c c c}
   \hline
   Nodal spacing & t=2.0 & t=4.0 &  t=6.0 & t=8.0 \\
   \hline
  h=L/10  & 0.9860e-004 & 0.9790e-004 & 0.9724e-004 & 0.9677e-004 \\
  h=L/20  & 0.2459e-004 & 0.2441e-004 & 0.2425e-004 & 0.2413e-004 \\
  h=L/40  & 0.6143e-005 & 0.6099e-005 & 0.6058e-005 & 0.6029e-005 \\
   \hline
 \end{tabular}
 \label{tab2}
  \end{table}

The square problem domain is modeled by $4\times4$, $8\times8$ and
$16\times16$ four-node square elements. The element interpolation
functions are:
\begin{eqnarray}
\begin{array}{c}
\displaystyle{N_1=(1-\xi)(1-\eta)/4,\,\,N_2=(1+\xi)(1-\eta)/4,\,\,\,\,\,\,\,\,\,\,\,\,\,\,\,\,\,\,\,\,\,\quad\quad\quad} \\
\displaystyle{N_3=(1+\xi)(1+\eta)/4,\,\,N_4=(1-\xi)(1+\eta)/4,\,\,\xi,
\eta \in [-1,\,1]}
\end{array}
\label{eq21}
\end{eqnarray}
%A visualization of the four-node square element for spatial
%discretization is shown in Figure \ref{fig3}.
and
\begin{eqnarray}
\textbf{C}^e=\frac{h^2}{36}\left[
\begin{array}{cccc}
     4 & 2 & 1 & 2\\
     2 & 4 & 2 & 1\\
     1 & 2 & 4 & 2\\
     2 & 1 & 2 & 4\\
\end{array}\right],\;
\textbf{K}^e=\frac{k}{6}\left[
\begin{array}{cccc}
     4 & -1 & -2 & -1\\
     -1 & 4 & -1 & -2\\
     -2 & -1 & 4 & -1\\
     -1 & -2 & -1 & 4\\
\end{array}\right].\;
\label{eq22}
\end{eqnarray}
%\begin{eqnarray}
%\textbf{K}^e=\frac{k}{6}\left[
%\begin{array}{cccc}
%     4 & -1 & -2 & -1\\
%     -1 & 4 & -1 & -2\\
%     -2 & -1 & 4 & -1\\
%     -1 & -2 & -1 & 4\\
%\end{array}\right].\;
%\label{eq23}
%\end{eqnarray}

\begin{figure}
\begin{center}
\includegraphics[width=0.8\textwidth]{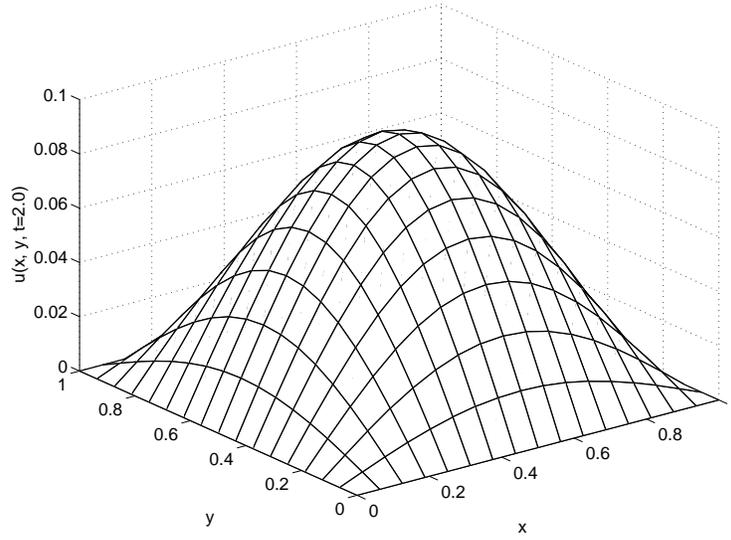}% Here is how to import EPS art
\end{center}\caption{The numerical result of two dimensional time-fractional diffusion model (\ref{eq20}) with $\gamma=0.8$ at
$t=2.0$. Diffusion coefficient $k=1/\pi^2$, space sizes $L=1.0$ and
nodal spacing $h=L/16$.} \label{fig4}
\end{figure}

Following the calculation steps (\ref{eq8})-(\ref{eq11}), Figure
\ref{fig4} plots the numerical solution for  $\gamma= 0.8$ at $t =
2$, the normalized errors for $\gamma= 0.8$ at $x = L/2$, $y = L/2$
and various values of $t$ are given in Table \ref{tab5}. The errors
drop with the nodal spacings. Indeed, the finite element method can
readily take coordinate-dependent and direction-dependent diffusion
coefficients into account.

%Following the calculation steps (\ref{eq8})-(\ref{eq11}), the
%numerical result of (\ref{eq20}) can be obtained. The concentration
%surface obtained by the time-fractional diffusion model (\ref{eq20})
%with $\gamma= 0.8$ at $t=2.0$ is shown in Figure \ref{fig4}, the
%normalized errors with different nodal spacings at different time
%constants are stated in Table \ref{tab5}. In this case, if the
%diffusion coefficients are functions of spatial variables, we only
%need to change the element matrix $\textbf{K}^e$ in the proposed
%method.

  \begin{table}
  \centering
   \caption{The normalized errors (Error$=|(u_{exact}(L/2, L/2, t)-u(L/2, L/2, t))/u_{exact}(L/2, L/2, t)|$) of the four-node square element. Space size $L=1.0$, nodal spacings $h=h_x=h_y$, diffusion
coefficient $k=1/\pi^2$  and time-fractional derivative order
$\gamma=0.8$ in (\ref{eq20}).}
\begin{tabular}{c  c  c  c c}
   \hline
   Space step & t=2.0 & t=4.0 &  t=6.0 & t=8.0 \\
   \hline
  h=L/4 & 6.5673e-002 & 6.1143e-002 & 5.7924e-002 & 5.6111e-002 \\
  h=L/8 & 1.6937e-002 & 1.5786e-002 & 1.4929e-002 & 1.4444e-002 \\
  h=L/16& 4.2664e-003 & 3.9778e-003 & 3.7602e-003 & 3.6368e-003 \\
   \hline
 \end{tabular}
 \label{tab5}
  \end{table}

\subsection{Time-fractional diffusion equation in a quarter of circular domain}
\begin{figure}
\begin{center}
\includegraphics[width=0.6\textwidth]{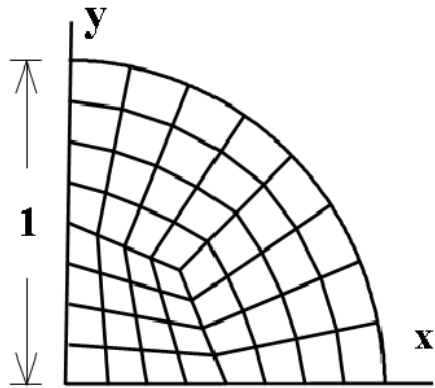}% Here is how to import EPS art
\end{center}\caption{The quadrant of a circle domain with unit radius
meshed into 48 elements.} \label{fig5}
\end{figure}

An important advantage of the finite element method over the finite
difference method is that the former can readily consider complex
spatial domains. In this example, the following problem defined over
a circular domain is considered:
\begin{eqnarray} \left\{
\begin{array}{c}
\displaystyle{\frac{d^\gamma u(x,y,t)}{d t^\gamma}= k (\frac{\partial^2 u(x,y,t)}{\partial x^2}+\frac{\partial^2 u(x,y,t)}{\partial y^2}),\,\,(x,y)\in \Omega,} \\
\displaystyle{\textbf{n}^T(\textbf{D}\nabla u)=0,\,\,x=0,\,y\neq 1,\,\,\text{and}\,\,y=0,\,0<x< 1,\, \,t>0},\\
\displaystyle{u(x,y,t)=J_0(1) E_\gamma (-t^\gamma),\,\,(x,y)\in \{(x,y)|x^2+y^2=1\}},\\
\displaystyle{ u(x,y,0)=J_0(\sqrt{x^2+y^2})},\,\,(x,y)\in \Omega
\cup \partial \Omega
\end{array}\right.\;
\label{eq24}
\end{eqnarray}
in which $\Omega= \{ (x, y) | x>0,\,\, y>0,\,\, x^2+y^2<1\}$.  If
$k=1$, the exact solution of (\ref{eq24}) is
$u_{exact}(x,y,t)=J_0(\sqrt{x^2+y^2}) E_\gamma (-t^\gamma)$, in
which $J_0$ represents the zeroth order Bessel function of the first
kind. For symmetry, we only need to model a quarter of the problem
domain and a typical mesh is depicted in Figure \ref{fig5}.

For the four-node element, the $(x,y)$ coordinates are also
interpolated with the functions given in (\ref{eq21}), i.e.
%Firstly, we employ the linear transformation functions to transform
%the actual four-node quadrilateral element into reference quadratic
%element. The new coordinates in the reference element can be
%expressed as
\begin{eqnarray}
\begin{array}{c}
\displaystyle{x^e(\xi,\eta)=\sum_{i=1}^{4} N_i x_i^e,\,\,y^e(\xi,\eta)=\sum_{i=1}^{4} N_i y_i^e} \\
\end{array}
\label{eq25}
\end{eqnarray}
in which ($x_i^e,\,y_i^e$) are the coordinates of the $i$-th element
nodes.
%in which $G_i^e$ is a linear transformation function in the form of
%(\ref{eq21}), $x_i^e,\,y_i^e$ are node coordinates in the actual
%quadrilateral element. Hereby the new coordinates are interpolated
%as
%\begin{eqnarray}
%\left[
%\begin{array}{c}
%     x(\xi,\eta)\\
%     y(\xi,\eta)
%\end{array}\right]
%=\left[
%\begin{array}{cccc}
%N_1 & N_2 & N_3 & N_4\\
%N_1 & N_2 & N_3 & N_4
%\end{array}\right]\left[
%\begin{array}{cc}
%x_1 & y_1\\
%x_2 & y_2\\
%x_3 & y_3\\
%x_4 & y_4
%\end{array}\right].
%\label{eq26}
%\end{eqnarray}
%Then the element matrices can be written as
%\begin{eqnarray}
%\left\{
%\begin{array}{c}
%\displaystyle{\int_\Omega \psi \frac{d^\gamma u}{d t^\gamma} d \Omega+ \int_\Omega \overrightarrow{grad}(\psi)\cdot \underline{D} \cdot\overrightarrow{grad}(u) d \Omega=0,} \\
%\displaystyle{u(\overrightarrow{X},0)=u_0(\overrightarrow{X})},
%\end{array}\right.\;
%\label{eq27}
%\end{eqnarray}

%\begin{eqnarray}
%\begin{array}{c}
%\displaystyle{\int_{\Omega^e} \{ N^e\} \cdot \langle N^e\rangle det(J) \frac{d^\gamma u^e}{d t^\gamma} d \Omega} \\
%\displaystyle{+ \int_{\Omega_e} \overrightarrow{grad}(\{N^e\})\cdot
%J^{-1} \cdot \underline{D} \cdot J^{-T} \cdot
%\overrightarrow{grad}(\langle N^e \rangle) det(J) \{u^e\} d
%\Omega=0},
%\end{array}.\;
%\label{eq28}
%\end{eqnarray}

%\begin{eqnarray}
%\begin{array}{c}
%\displaystyle{[\textbf{C}^e]\{\frac{d^\gamma u^e}{d
%t^\gamma}\}+[\textbf{K}^e] \{ u^e\}= 0},
%\end{array}
%\label{eq29}
%\end{eqnarray}
%where
\begin{eqnarray}
\left\{
\begin{array}{c}
\displaystyle{\textbf{C}^e=\int_{-1}^1 \int_{-1}^1  (\textbf{N}^e)^T   \textbf{N}^e det(\textbf{J}) d \xi d\eta,} \\
\displaystyle{\textbf{K}^e= \int_{-1}^1 \int_{-1}^1 (\nabla
\textbf{N}^e)^T \textbf{J}^{-1} \textbf{D}  \textbf{J}^{-T}  (\nabla
\textbf{N}^e) det(\textbf{J}) d \xi d\eta}.
\end{array}\right.\;
\label{eq30}
\end{eqnarray}
In the above expression,
\begin{eqnarray}
\nabla=\left\{
\begin{array}{c}
     \partial/\partial \xi\\
       \partial/\partial \eta
\end{array}\right \}\;
\text{and} \,\,\,\,\textbf{J}=\left[
\begin{array}{cc}
     \partial x^e/\partial \xi & \partial y^e/\partial \xi\\
     \partial x^e/\partial \eta & \partial y^e/\partial \eta
\end{array}\right].\;
\label{eq14}
\end{eqnarray}
%$\textbf{J}$ is the Jacobian matrix, $det(\textbf{J})$ is the
%Jacobian matrix determinant, $\textbf{D}$
%is the diffusion coefficient matrix.\\
The matrices $C^e$ and $K^e$ are computed by the second order
Gauss-Legendre rule. In our computations, a quarter circle is
partitioned into 3, 48 and 217 elements, the corresponding numerical
results at some selected spatial locations are listed in Table
\ref{tab6} for $\gamma=0.8$. Table \ref{tab6} indicates that the
proposed method is capable of delivering accurate solution to
anomalous diffusion problem (\ref{eq24}) and the accuracy can be
improved by employing more elements in modeling the computational
domain.
%\begin{eqnarray}
%[J]=\left[
%\begin{array}{cc}
%     \frac{\partial x}{\partial \xi} & \frac{\partial x}{\partial \eta}\\
%     \frac{\partial y}{\partial \xi} & \frac{\partial y}{\partial \eta}
%\end{array}\right],\;
%\label{eq31}
%\end{eqnarray}

  \begin{table}
  \centering
   \caption{The numerical results by different numbers of elements, In the computation, $k(x)=k(y)=1$, the exact solution of (\ref{eq20}) is
$u_{exact}(x,y,t)=J_0(\sqrt{x^2+y^2}) E_\gamma
(-t^\gamma),\,\,\gamma=0.8$.}
\begin{tabular}{c  c  c  c c}
   \hline
   Coordinates & 3 Elements & 48 Elements &  217 Elements & Exact solution \\
   \hline
  (0,0)              & 0.37770 & 0.38638 & 0.38681 & 0.38695 \\
  (0.35355, 0.35355) & 0.34055 & 0.36233 & 0.36292 & 0.36314 \\
  (0.21339, 0.21339) & ---     & 0.37760 & 0.37804 & 0.37819 \\
  (0.42678, 0.17678) & ---     & 0.36603 & 0.36644 & 0.36658 \\
  (0.67533, 0.27973) & ---     & 0.33659 & 0.33687 & 0.33696 \\
  (0.53033, 0.53033) & ---     & 0.33404 & 0.33432 & 0.33442 \\
  (0.27973, 0.67533) & ---     & 0.33659 & 0.33687 & 0.33696 \\
  (0.17678, 0.42678) & ---     & 0.36603 & 0.36644 & 0.36658 \\
   \hline
 \end{tabular}
 \label{tab6}
  \end{table}

\section{Application}

\begin{figure}
\begin{center}
\includegraphics[width=0.6\textwidth]{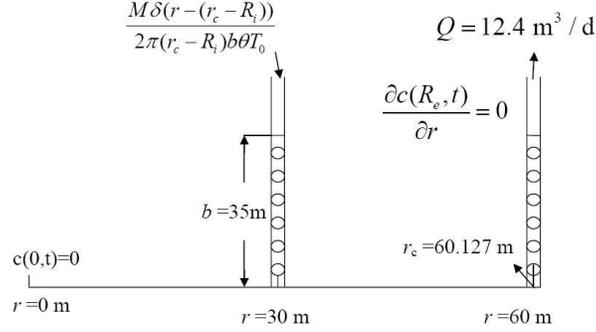}% Here is how to import EPS art
\end{center}\caption{A schematic diagram of experiment.} \label{fig10}
\end{figure}

To investigate the efficiency and applicability of the proposed
semi-analytical method in solving real-world problems, it is
employed to solve the problem of tracer solute transport in an
aquifer. The experiment was conducted using a test aquifer in Nevada
and the schematic diagram is shown in Figure \ref{fig10}
\cite{Benson2004}. Bromide of quantity $M=20.81$ kg was used as a
nonsorbing tracer solute and introduced to the injection well for a
period of $T_0=3.54$ days at an average concentration of $3.77$
kg/m$^3$. A reference point, the injection well and extraction well
are located at $r = 0$ m , 30 m ($R_i$) and 60 m ($R_e$)  along the
downstream direction of the underground water flow. The radius of
the extraction well is $0.127$ m, the center of extraction well is
at $r_c=60.127$ m, the pumping rate is $Q=12.4$ m$^3$/d. The solute
concentration in the extraction well had been monitored for about
321 days and the screened interval was $b=35$ m . More detailed
description of the experiment can be found in
\cite{Benson2004,Meerschaert2004,Reimus2003,Pohll1999,Reimus2003b}.
%The distance from injection well to the extraction well is $30 m$,
%and the radius of extraction well is $r_e=0.127 m$. The extraction
%well screened interval is $b=35 m$, the outer and inner boundaries
%are $R_O =-60.127m$ and $R_I = -0.127 m$, the injection well and
%extraction well are at $R_L = -30.127 m$ and $R_I = -0.127 m$
%\cite{Benson2004,Meerschaert2004,Reimus2003}. The detailed
%description of this experiment is stated in
%\cite{Pohll1999,Reimus2003}.

%Assuming the medium is radially homogeneous, Meerschaert and
% Benson et al. have established spatial fractional equation and
%temporal-spatial fractional equation as the governing equations to
%investigate the tracer solute transport in the fractured granite
%aquifer \cite{Benson2004,Meerschaert2004}.
Since $T_0$ is short compared with the total time range ($\sim$ 321
days) of measurement, the following radial initial-boundary value
problem for the solute transport in the fractured granite aquifer is
established in terms of the solute concentration $u$ as:
\begin{eqnarray}
\left\{
\begin{array}{c}
\displaystyle{\frac{d^\gamma u(r,t)}{d t^\gamma} =-\frac{\upsilon_0}{r_c-r} \frac{\partial u(r,t)}{\partial r}+ \frac{1}{r_c-r} \frac{\partial}{\partial r} (d_0 \frac{\partial u(r,t)}{\partial r}),\,\,r\in (0,R_e),} \\
\displaystyle{u(0,t)=0,\,\, \,\, \frac{\partial u(R_e,t)}{\partial r}=0,\,\,t>0,\,\,}\\
u(r,0)=f(r),\,\,r\in [0,R_e]
\end{array}\right.\;
\label{eq32}
\end{eqnarray}
where $\upsilon_0$ is the convective coefficient, $d_0$ is the
dispersion coefficient, $\upsilon_0=a d_0$ and $a$ is the
dispersivity. Moreover $\upsilon_0/(r_c-r)$ and $d_0/(r_c-r)$ have
the units of $[L/T^\gamma]$ and $[L^2/T^\gamma]$ which represent the
nonlocal aquifer properties \cite{Zhang2009}. The initial value is
normalized as $f(r)=M \delta(r- (r_c-R_i))/(2\pi (r_c-R_i) b \theta
T_0)$, $\theta$ is the hydraulic parameter. The boundary conditions
imply that the solute cannot reach $r=0$ by upstream dispersion and
the solute moves by advection at $r=R_e$ which gives the wall of the
extraction well \cite{Benson2004}.
%For computation convenience, we
%firstly transform the governing equation into the following form:
%\begin{eqnarray}
%\begin{array}{c}
%\displaystyle{(r_c-r) \frac{d^\gamma u(r,t)}{d t^\gamma}=-\upsilon_0
%\frac{\partial u(r,t)}{\partial r}+ d_0 \frac{\partial^2
%u(r,t)}{\partial r^2}.}
%\end{array}
%\label{eq33}
%\end{eqnarray}
%
%It can be proved that the boundary conditions and initial condition
%are satisfied. Then we can obtain the weak integral formulation of
%equation (\ref{eq33}).

In order to obtain a high accurate numerical approximation, the
quadratic element is adopted, the element matrices $C^e$ and $K^e$
are computed by the second order Gauss-Legendre rule.

%the corresponding element matrices $\textbf{C}_i^e$ ($i$ is the
%element index) and $\textbf{K}^e$ are:
%\begin{eqnarray}
%\textbf{C}_i^e=\frac{h}{60}\left[
%\begin{array}{ccc}
%     \displaystyle{h + 8 r_i} & \displaystyle{4 r_i} & \displaystyle{-h - 2 r_i} \\
%     \displaystyle{4 r_i} & \displaystyle{16 h + 32 r_i } & \displaystyle{4 h + 4 r_i }\\
%    \displaystyle{-h - 2 r_i} & \displaystyle{4 h + 4 r_i} & \displaystyle{7 h + 8
%     r_i}
%\end{array}\right]\;
%\label{eq34}
%\end{eqnarray}

%\begin{eqnarray}
%\textbf{C}^e=\frac{h}{30}\left[
%\begin{array}{ccc}
%     4 & 2 & -1\\
%     2 & 16 & 2\\
%     -1 & 2 & 4
%\end{array}\right],\;
%\label{eq16}
%\end{eqnarray}
%
%
%\begin{eqnarray}
%\textbf{K}^e=\frac{\upsilon_0}{6}\left[
%\begin{array}{ccc}
%     -3 & 4 & -1\\
%     -4 & 0 & 4\\
%     1  & -4 & 3
%\end{array}\right]\;
%+\frac{d_0}{3 h}\left[
%\begin{array}{ccc}
%     7 & -8 & 1\\
%     -8 & 16 & -8\\
%     1 & -8 & 7
%\end{array}\right]\;
%\label{eq35}
%\end{eqnarray}
%in which $r_i$ denotes the coordinate of the left hand node in the
%$i$-th element.

\begin{figure}
\begin{center}
\includegraphics[width=0.8\textwidth]{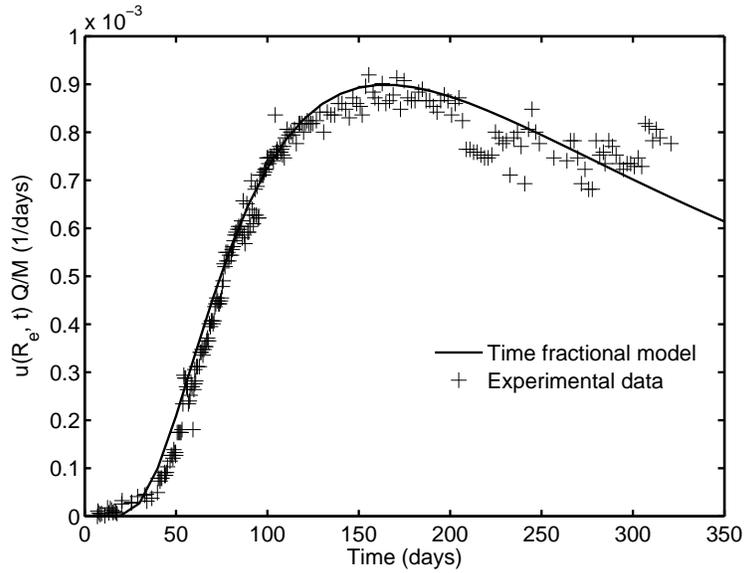}% Here is how to import EPS art
\end{center}\caption{The numerical approximation of the time-fractional
radial flow advection-dispersion equation model (\ref{eq24}) with
$\gamma=0.92$ at $r=R_e$. In this numerical simulation, convective
coefficient $\upsilon_0=0.0564/\theta,\,\, \theta=0.023$, the
dispersion coefficient $d_0=a \upsilon_0,\,\, a=6.8$, nodal spacing
$h=3$ and time step is $\Delta t=10$.} \label{fig6}
\end{figure}

\begin{figure}
\begin{center}
\includegraphics[width=0.8\textwidth]{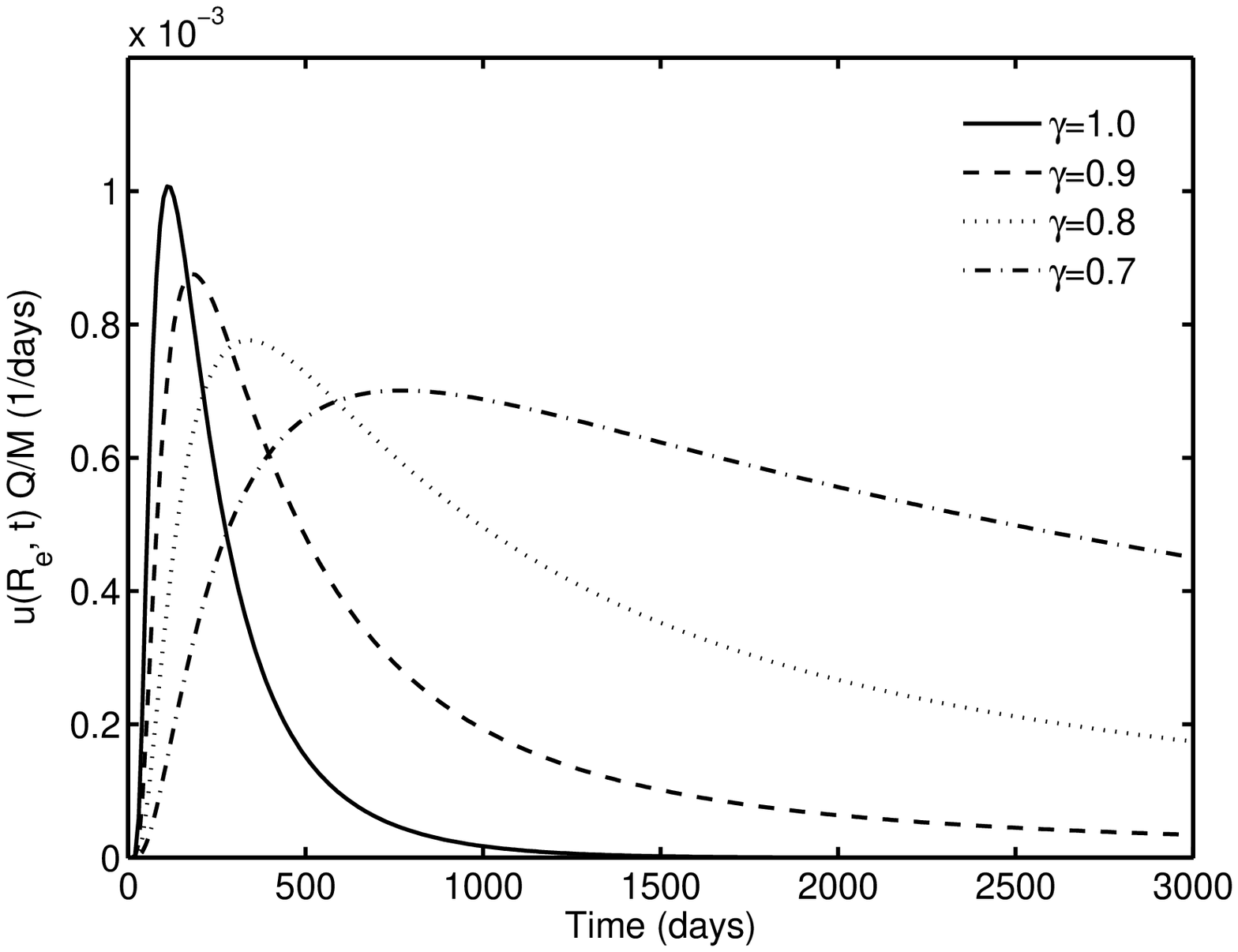}% Here is how to import EPS art
\end{center}\caption{The numerical approximation of the time-fractional
radial flow advection-dispersion equation model (\ref{eq24}) with
different time-fractional derivative values $\gamma$ at $r=R_e$. In
this numerical simulation, convective coefficient
$\upsilon_0=0.0564/\theta,\,\, \theta=0.023$, dispersion coefficient
$d_0=a \upsilon_0,\,\, a=6.8$, nodal spacing $h=3$ and time step is
$\Delta t=10$. The curve with $\gamma=1.0$ corresponding to Fickian
dispersion.} \label{fig7}
\end{figure}

A comparison of the numerical predictions and the experimental data
is shown in Figure \ref{fig6} and the heavy-tail feature
characterized by the time-fractional model (\ref{eq32}) with
different derivative orders is shown in Figure \ref{fig7}. It can be
observed from Figure \ref{fig6} that the numerical result offers a
good fit to most of the experimental data. Due to the subdiffusion
behavior in the aquifer matrix and immobile water, in the
experimental result, the concentration of bromide exhibits a rather
slow decay in the late time. Figure \ref{fig7} confirms that the
time-fractional radial flow model (\ref{eq32}) captures the
long-time behavior with heavy-tail. Figure \ref{fig7} also
illustrates that the heavy-tail feature becomes more remarkable with
the decreasing of the time-fractional derivative order $\gamma$.
Hence, in this model (\ref{eq32}), the time-fractional derivative
order $\gamma$ is a indicator of the non-Fickian transport caused by
the complex structure of the fractured aquifer.
%The fractional radial flow model captures the early breakthrough of
%bromide at the extraction well, as well as the late-time tailing.e
%the Long time behavior with heavy tail Time fractional model: Long
%time behavior with heavy tail; Spatial fractional model: anomalously
%early arrivals, early arrival This is accompanied by an overall
%increase in the width of the breakthrough curve. \textbf{The effects
%of the fractional space and time derivatives are, for the most part,
%independent.} it is most important to have reliable models of the
%leading edge of the breakthrough curve. "so the applicability of the
%more general mobile/ immobile, CTRW, or time-fractional models (4)
%is well established."

\section{Discussions}

By using the finite element to discretize the spatial domain, the
fractional diffusion equations can be reduced to approximate
fractional relaxation equations which possess analytical solutions.
The semi-analytical method can not only compute time-fractional
diffusion equations in long-time range at low computational cost but
also deliver accurate numerical predictions for complex and large
spatial problem domains. The accuracy in spatial domain can be
improved by using more elements, high-order elements or elements
based on advanced finite element formulations. Since the exact
solution is used in time domain, the stability and convergence
conditions of the proposed method can be easily satisfied. It can be
said that the proposed method is more robust than previous ones.

The main restriction for the proposed method is that the weak forms
of time diffusion equations can be transformed into the following
form:
\begin{eqnarray}
\begin{array}{c}
\displaystyle{\textbf{C}\frac{d^\gamma }{d
t^\gamma}\textbf{u}+\textbf{K} \textbf{u}= 0}.
\end{array}
\label{eq36}
\end{eqnarray}
In cases that the source term, physical parameters and/or boundary
conditions are only weak function(s) of time, a multiple time step
method can be used.

%From the observation of (\ref{eq0}), we know that the considered
%equation does not contain source terms. However, it should be
%pointed out that some source term cases can also be solved by the
%proposed method. For example, if the source term is $f(x)=P e^x$
%($P$ is a constant) in one dimension, then we can let
%$\upsilon=u+f(x)$ and substitute it into (\ref{eq0}), hereby it can
%be solved by the proposed semi-analytical finite element method
%under suitable boundary conditions and system parameters.

%
%\textbf{Need to improve}
%\textbf{Question}: In which condition of $q(u,x,t)$, the following
%equation can be solved by the current method?
%\begin{eqnarray}
%\left\{
%\begin{array}{c}
%\displaystyle{_0^CD_t^\gamma u(x,t)=d(x,t)(\frac{\partial^2 u(x,t)}{\partial x^2}+\frac{\partial^2 u(x,t)}{\partial y^2}),\,\,(x,y)\in \Omega,\,\,t>0,} \\
%\displaystyle{\frac{\partial u(x,y,t)}{\partial n}=q(u,x,t),\,\,(x,y)\in \partial \Omega,}\\
%u(x,y,0)=sin(x\pi/L_x)sin(y\pi/L_y),
%\end{array}\right.\;
%\label{eq621}
%\end{eqnarray}

\section{Concluding remarks}

From formulations and examples presented, it is clear that a class
of time-fractional diffusion equations can be easily computed and
the heavy-tail feature can be accurately characterized by the new
method. Our future research work will focus on advanced finite
element formulations, such as hybrid element \cite{Sze2008}, to
compute temporal-spatial fractional diffusion equations which
characterize more complex contaminant transport problems.

%The proposed scheme can also be extend to solve anomalous problems
%characterized by time fractional derivatives in other physical
%fields, such as heat transfer and wave dissipation.

\section*{Acknowledgement}
The first author thanks Prof. G. Pohll and Prof. M. M. Meerschaert
for providing the experimental data, Dr. Q. H. Zhang for valuable
discussions on finite element programming. The work described in
this paper was supported by the National Basic Research Program of
China (973 Project No. 2010CB832702), the R$\&$D Special Fund for
Public Welfare Industry (Hydrodynamics, Project No. 201101014) and
the Opening Fund of the State Key Laboratory of Structural Analysis
for Industrial Equipment (Project No. GZ0902).

\newpage

%\section*{Figure introduction}
%
%Fig.1 The diffusion behaviors of time dependent VODO model and
%constant order model at $x=0.6$. The blue, green and black lines
%note the concentration curve with $\alpha=0.6$, $0.7$ and $0.8$; the
%red line represents the concentration evolution
%curve with VO function is (\ref{eq7}).\\
%
%Fig.2 The comparison of two types of space dependent VO models and
%CO model  at $t=10$. the green and black lines note the
%concentration curves with $\alpha=0.5$ and $\alpha=0.6$; the red and
%blue lines represent the concentration curve with VO functions are
%(\ref{eq9}) and (\ref{eq10}).\\
%
%Fig.3 The diffusion behaviors of concentration dependent VODO model
%and CO model at $x=0.5$. The blue, green lines note the
%concentration curve with $\alpha=0.5$ and $\alpha=0.6$; the red line
%represents the concentration evolution curve with VO function is
%(\ref{eq12}).

%\begin{figure}
%\begin{center}
%\includegraphics[width=0.8\textwidth]{element}% Here is how to import EPS art
%\end{center}\caption{The mesh grid of spatial domain in (\ref{eq20}) and the
%sequence of nodes.} \label{fig3}
%\end{figure}

\begin{thebibliography}{00}

\bibitem{Dagan1987}
G. Dagan. Theory of solute transport by groundwater.
\newblock Ann Rev Fluid Mech
\newblock 1987; 19: 183-215.

\bibitem{LaBolle2001}
E. M. LaBolle, G. E. Fogg. Role of molecular diffusion in
contaminant migration and recovery in an alluvial aquifer system.
\newblock Transport Porous Med
\newblock 2001; 42: 155-179.

\bibitem{Zhang2009}
Y. Zhang, D. A. Benson, D. M. Reeves. Time and space nonlocalities
underlying fractional-derivative models: Distinction and literature
review of field applications.
\newblock Adv Water Resour
\newblock  2009; 32: 561-581.

\bibitem{Metzler2000}
R. Metzler, J. Klafter. The random walk's guide to anomalous
diffusion: a fractional dynamics approach.
\newblock Phys Rep
\newblock  2000; 339: 1-77.

\bibitem{Huang2008}
Q. Z. Huang, G. H. Huang, H. B. Zhan. A finite element solution for
the fractional advection-dispersion equation.
\newblock Adv Water Resour
\newblock 2008; 31: 1578-1589.

\bibitem{Berkowitz2006}
B. Berkowitz, A. Cortis, M. Dentz,  H. Scher. Modeling non-Fickian
transport in geological formations as a continuous time random walk.
\newblock Rev Geophys
\newblock  2006; 44(2): RG2003.

\bibitem{Berkowitz1995}
B. Berkowitz, H. Scher. On characterization of anomalous dispersion
in porous media.
\newblock Water Resour Res
\newblock  1995; 31: 1461-1466.

\bibitem{Zhang2007}
X. X. Zhang, M. Lv, J. W. Crawford, I. M. Young. The impact of
boundary on the fractional advection¨Cdispersion equation for solute
transport in soil: Defining the fractional dispersive flux with the
Caputo derivatives.
\newblock Adv Water Resour
\newblock  2007; 30: 1205-1217.

\bibitem{Sun2009}
H. G. Sun, W. Chen, Y. Q. Chen. Variable-order fractional
differential operator in anomalous diffusion modeling.
\newblock Phys A
\newblock  2009; 388: 4586-4592.

\bibitem{Benson2000}
D. A. Benson, S. W. Wheatcraft, M. M. Meerschaert. Application of a
fractional advection-dispersion equation.
\newblock Water Resour Res
\newblock 2000; 36(6): 1403-1412.

\bibitem{Seymour2007}
J. D. Seymour,  J. P. Gage, S. L. Codd, R. Gerlach. Magnetic
resonance microscopy of biofouling induced scale dependent transport
in porous media.
\newblock Adv Water Resour
\newblock  2007; 30(6-7): 1408-1420.

\bibitem{Meerschaert2001}
M. M. Meerschaert, D. A.  Benson, B. Baeumer. Operator
L$\acute{e}$vy motion and multiscaling anomalous diffusion.
\newblock Phys Rev E
\newblock 2001; 63: 021112.

\bibitem{Benson2004}
D. A. Benson, C. Tadjeran, M. M. Meerschaert, I. Farnham, G. Pohll.
Radial fractional-order dispersion through fractured rock.
\newblock Water Resour Res
\newblock 2004; 40: W12416.

\bibitem{Sokolov2005}
I. M. Sokolov, J. Klafter. From diffusion to anomalous diffusion: A
century after Einstein's Brownian motion.
\newblock Chaos
\newblock  2005; 15: 026103.

\bibitem{Zaslavsky2002}
G. M. Zaslavsky. Chaos, fractional kinetics, and anomalous
transport.
\newblock Phys Rep
\newblock  2002; 371 (6): 461-580.

\bibitem{Magin2008}
R. L. Magin, O. Abdullah, D. Baleanu, et al. Anomalous diffusion
expressed through fractional order differential operators in the
Bloch-Torrey equation.
\newblock  J Magn Reson
\newblock  2008; 190 (2): 255-270.

\bibitem{Podlubny2009}
I. Podlubny, A. Chechkin, T. Skovranek, Y. Q. Chen,  B. M. Vinagre
Jara. Matrix approach to discrete fractional calculus II: Partial
fractional differential equations.
\newblock J Comput Phys
\newblock  2009; 228: 3137-3153.

\bibitem{Roop2006}
J. P. Roop. Computational aspects of FEM approximation of fractional
advection dispersion equations on bounded domains in $\mathbb{R}^2$.
\newblock J Comput Appl Math
\newblock  2006; 193: 243-268.

\bibitem{Li2011}
C. Li, A. Chen, J. J. Ye. Numerical approaches to fractional
calculus and fractional ordinary differential equation.
\newblock J Comput Phys
\newblock  2011; 230(9): 3352-3368.

\bibitem{Podlubny1999}
I. Podlubny.  Fractional differential equation. San Diego, Academic
press, 1999. 50-78.

\bibitem{Ford2001}
N. J. Ford, A. C. Simpson. The numerical solution of fractional
differential equations: speed versus accuracy.
\newblock Numer Algorithms
\newblock  2001; 26: 336-346.

\bibitem{Radu2011}
F. A. Radu, N. Suciu, J. Hoffmann, A. Vogel, O. Kolditz, C. H. Park,
S. Attinger. Accuracy of numerical simulations of contaminant
transport in heterogeneous aquifers: A comparative study.
\newblock Adv Water Resour
\newblock  2011; 34: 47-61.

\bibitem{Zienkiewicz2000}
O. C. Zienkiewicz, R. L. Taylor. The finite element method: Volume 3
Fluid Dynamics fifth edition. Oxford, Butterworth-Heinemann, 2000.

\bibitem{Lewis1996}
R. W. Lewis, K. Morgan, H. R. Thomas, K. N. Seetharamu. The finite
element method in heat transfer analysis. New York, John Wiley \&
Sons, 1996.

\bibitem{Bergheau2008}
J. M. Bergheau, R. Fortunier. Finite element simulation of heat
transfer. London, John Wiley \& Sons, 2008.

\bibitem{Smith2004}
I. M. Smith, D. V. Griffiths.  Programming the finite element method
(4th edition) . New York, John Wiley \& Sons Ltd, 2004.

\bibitem{Yuste2006}
S. B. Yuste. Weighted average finite difference methods for
fractional diffusion equations.
\newblock J Comput Phys
\newblock  2006; 216: 264-274.

\bibitem{Chen2007}
C. M. Chen, F. Liu, I. Turner, V. Anh. A Fourier method for the
fractional diffusion equation describing sub-diffusion.
\newblock J Comput Phys
\newblock  2007; 227: 886-897.

\bibitem{Fix2004}
G. J. Fix, J. P. Roop. Least squares finite-element solution of a
fractional order two-point boundary value problem.
\newblock Comput \& Math Appl
\newblock  2004; 48 (7-8): 1017-1033.

\bibitem{Deng2008}
W. H. Deng. Finte element method for the space and time fractional
Fokker-Planck equation.
\newblock SIAM J Numer Anal
\newblock 2008; 47(1):204-226.

\bibitem{Zhang2010}
H. Zhang, F. Liu, V. Anh. Galerkin finite element approximation of
symmetric space-fractional partial differential equations.
\newblock Appl Math Comput
\newblock 2010; 217(6):2534-2545.

\bibitem{Zheng2010}
Y. Y. Zheng, C. P. Li, Z. G. Zhao. A note on the finite element
method for the space-fractional advection diffusion equation.
\newblock Comput \& Math Appl
\newblock 2010; 59: 1718-1726.

\bibitem{Zheng2010b}
Y. Y. Zheng, C. P. Li, Z. G. Zhao. A fully discrete discontinuous
Galerkin method for nonlinear fractional Fokker-Planck equation.
\newblock Math Probl Engng
\newblock 2010; doi:10.1155/2010/279038.

\bibitem{Li2011}
C. P. Li, Z. G. Zhao, Y. Q. Chen. Numerical approximation of
nonlinear fractional differential equuations with subdiffusion and
superdiffusion.
\newblock Comput \& Math Appl
\newblock 2011; doi:10.1016/j.camwa.2011.02.045.

\bibitem{Samko1993}
S. G. Samko, A. A. Kilbas, O. I. Marichev.  Fractional integrals and
derivatives: theory and applications. Gordon and Breach, Taylor \&
Francis Ltd, 1993.

\bibitem{Kumar2006}
P. Kumar, O. P. Agrawal. An approximate method for numerical
solution of fractional differential equations.
\newblock Signal Processing
\newblock 2006; 86 (10), 2602-2610.

\bibitem{Guymon1970}
G. L. Guymon. A finite element solution of the one-dimensional
diffusion-convection equation.
\newblock Water Resour Res
\newblock  1970; 6(1): 204-210.

\bibitem{Mainardi1996}
F. Mainardi. Fractional relaxation-oscillation and fractional
diffusion-wave phenomena.
\newblock Chaos, Solitons \& Fractals
\newblock 1996; 7(9): 1461-1477.

\bibitem{Podlubny2009a}
I. Podlubny. Mittag-Leffler function.
\newblock http://www.mathworks.de /matlabcentral/fileexchange/8738-mittag-leffler-function
\newblock 2009.

\bibitem{Chen2008}
Y. Q. Chen. Generalized Mittag-Leffler function.
\newblock http://www.mathworks.de/matlabcentral/fileexchange/20849-generalized-mittag-leffler-function
\newblock 2008.

\bibitem{Su2009}
L. J. Su, W. Q. Wang, Z. X. Yang. Finite difference approximations
for the fractional advection-diffusion equation.
\newblock Phys Lett A
\newblock 2009; 373: 4405-4408.

\bibitem{Meerschaert2004}
M. M. Meerschaert, C. Tadjeran. Finite difference approximations for
fractional advection-dispersion flow equations.
\newblock J Comput Appl Math
\newblock 2004; 172: 65-77.

\bibitem{Reimus2003}
P. Reimus, G. Pohll, T. Mihevc, J. Chapman, M. Haga, B. Lyles, S.
Kosinski, R. Niswonger, P. Sanders. Testing and parameterizing a
conceptual model for solute transport in a fractured granite using
multiple tracers in a forced-gradient test.
\newblock Water Resour Res
\newblock 2003; 39: 1356-1370.

\bibitem{Pohll1999}
G. Pohll, A. E. Hassan, J. B. Chapman, C. Papelis, R. Andricevic.
Modeling ground water flow and radioactive transport in a fractured
aquifer.
\newblock Ground Water
\newblock 1999; 37(5): 770-784.

\bibitem{Reimus2003b}
P. W. Reimus, M. J. Haga, A. I. Adams, T. J. Callahan, H. J. Turin,
D. A. Coun. Testing and parameterizing a conceptual solute transport
model in saturated fractured tuff using sorbing and nonsorbing
tracers in cross-hole tracer tests.
\newblock J Contam Hydrol
\newblock 2003; 62-63: 613-636.

\bibitem{Sze2008}
K. Y. Sze, Y. K. Cheung. A hybrid-Trefftz finite element model for
Helmholtz problem.
\newblock Commun Numer Meth Engng
\newblock 2008; 24: 2047-2060.
%\bibitem{Zhang2009}
%Zhang Y, Benson DA, Reeves DM. Time and space nonlocalities
%underlying fractional-derivative models: Distinction and literature
%review of field applications.
%\newblock \emph {Adv Water Resour},
%\newblock  2009; 32: 561-581.
%
%\bibitem{Metzler2000}
%Metzler R, Klafter J. The random walk's guide to anomalous
%diffusion: a fractional dynamics approach.
%\newblock \emph {Phys Rep},
%\newblock  2000; 339: 1-77.
%
%\bibitem{Huang2008}
%Huang QZ, Huang GH, Zhan HB. A finite element solution for the
%fractional advection-dispersion equation.
%\newblock \emph {Adv Water Resour}
%\newblock 2008; 31: 1578-1589.
%
%\bibitem{Berkowitz2006}
%Berkowitz B, Cortis A, Dentz M,  Scher H. Modeling non-Fickian
%transport in geological formations as a continuous time random walk.
%\newblock \emph {Rev Geophys},
%\newblock  2006; 44(2): RG2003.
%
%\bibitem{Berkowitz1995}
%Berkowitz B, Scher H. On characterization of anomalous dispersion in
%porous media.
%\newblock \emph {Water Resour Res},
%\newblock  1995; 31: 1461-1466.
%
%\bibitem{Zhang2007}
%Zhang XX, Lv M, Crawford JW, Young IM. The impact of boundary on the
%fractional advection¨Cdispersion equation for solute transport in
%soil: Defining the fractional dispersive flux with the Caputo
%derivatives.
%\newblock \emph {Adv Water Resour},
%\newblock  2007; 30: 1205-1217.
%
%\bibitem{Sun2009}
%Sun HG, Chen W, Chen YQ. Variable-order differential operator in
%anomalous diffusion modeling.
%\newblock \emph {Phys A},
%\newblock  2009; 388: 4586-4592.
%
%\bibitem{Benson2000}
%Benson DA, S. W. Wheatcraft, M. M. Meerschaert, Application of a
%fractional advection-dispersion equation.
%\newblock \emph {Water Resour. Res.}
%\newblock 2000, 36(6): 1403-1412.
%
%\bibitem{Seymour2007}
%J. D. Seymour, J. P. Gage, S. L. Codd, R. Gerlach. Magnetic
%resonance microscopy of biofouling induced scale dependent transport
%in porous media.
%\newblock \emph {Advances in Water Resources},
%\newblock  2007, 30(6-7): 1408-1420.
%
%\bibitem{Meerschaert2001}
%M. M. Meerschaert, D. A. Benson, B. Baeumer, Operator L$\acute{e}$vy
%motion and multiscaling anomalous diffusion.
%\newblock \emph {Phys. Rev. E}
%\newblock 2001, 63: 021112.
%
%\bibitem{Benson2004}
%D. A. Benson, C. Tadjeran, M. M. Meerschaert, I. Farnham, G. Pohll.
%Radial fractional-order dispersion through fractured rock.
%\newblock \emph {Water Resource Research}
%\newblock 2004, 40:W12416.
%
%\bibitem{Sokolov2005}
%I. M. Sokolov and J. Klafter. From diffusion to anomalous diffusion:
%A century after Einstein's Brownian motion.
%\newblock \emph {Chaos},
%\newblock  2005, 15: 026103.
%
%\bibitem{Zaslavsky2002}
%G. M. Zaslavsky. Chaos, fractional kinetics, and anomalous
%transport.
%\newblock \emph {Phys. Rep.},
%\newblock  2002, 371 (6): 461-580.
%
%\bibitem{Podlubny2009}
%I. Podlubny, A. Chechkin, T. Skovranek, Y. Q. Chen, B. M. Vinagre
%Jara. Matrix approach to discrete fractional calculus II: Partial
%fractional differential equations.
%\newblock \emph {J. Comput. Phys.},
%\newblock  2009, 228: 3137-3153.
%
%\bibitem{Roop2006}
%J. P. Roop. Computational aspects of FEM approximation of fractional
%advection dispersion equations on bounded domains in $\mathbb{R}^2$.
%\newblock \emph {Journal of Computational and Applied Mathematics},
%\newblock  2006, 193: 243-268.
%
%\bibitem{Radu2011}
%F.A. Radu, N. Suciu, J. Hoffmann, A. Vogel, O. Kolditz, C.-H. Park,
%S. Attinger. Accuracy of numerical simulations of contaminant
%transport in heterogeneous aquifers: A comparative study.
%\newblock \emph {Advances in Water Resources},
%\newblock  2011, 34: 47-61.
%
%\bibitem{Zienkiewicz2000}
%O. C. Zienkiewicz, R. L. Taylor. \emph{The finite element method:
%Volume 3 Fluid Dynamics fifth edition}. Oxford,
%Butterworth-Heinemann, 2000.
%
%\bibitem{Lewis1996}
%R. W. Lewis, K. Morgan, H. R. Thomas, K. N. Seetharamu. \emph{The
%finite element method in heat transfer analysis}. New York, John
%Wiley \& Sons, 1996.
%
%\bibitem{Bergheau2008}
%J.-M. Bergheau, R. Fortunier. \emph{Finite element simulation of
%heat transfer}. London, John Wiley \& Sons, 2008.
%
%\bibitem{Yuste2006}
%S. B. Yuste, Weighted average finite difference methods for
%fractional diffusion equations.
%\newblock \emph {J. Comput. Phys.},
%\newblock  2006, 216: 264-274.
%
%\bibitem{Chen2007}
%C. M. Chen, F. Liu, I. Turner, V. Anh, A Fourier method for the
%fractional diffusion equation describing sub-diffusion.
%\newblock \emph {J. Comput. Phys.},
%\newblock  2007, 227: 886-897.
%
%\bibitem{Fix2004}
%G. J. Fix, J. P. Roop. Least squares finite-element solution of a
%fractional order two-point boundary value problem.
%\newblock \emph {Computers} \& \emph{Mathematics with Applications},
%\newblock  2004, 48 (7-8): 1017-1033.
%
%\bibitem{Deng2008}
%W. H. Deng. Finte element method for the space and time fractional
%Fokker-Planck equation.
%\newblock \emph {SIAM J. Numer. Anal.}
%\newblock 2008, 47(1):204-226.
%
%\bibitem{Zhang2010}
%H. Zhang, F. Liu, V. Anh. Galerkin finite element approximation of
%symmetric space-fractional partial differential equations.
%\newblock \emph {Applied Mathematics and Computation}
%\newblock 2010, 217(6):2534-2545.
%
%\bibitem{Zheng2010}
%Y. Y. Zheng, C. P. Li, Z. G. Zhao. A note on the finite element
%method for the space-fractional advection diffusion equation.
%\newblock \emph {Computers and Mathematics with Applications}
%\newblock 2010, 59: 1718-1726.
%
%\bibitem{Samko1993}
%S. G. Samko, A.A. Kilbas and O.I. Marichev.  \emph{Fractional
%integrals and derivatives: theory and applications }. Gordon and
%Breach, Taylor \& Francis Ltd, 1993.
%
%\bibitem{Kumar2006}
%P. Kumar, O. P. Agrawal. An approximate method for numerical
%solution of fractional differential equations.
%\newblock \emph {Signal Processing}
%\newblock 2006, 86 (10), 2602-2610.
%
%\bibitem{Guymon1970}
%G. L. Guymon. A finite element solution of the one-dimensional
%diffusion-convection equation.
%\newblock \emph {Water Resource Research},
%\newblock  1970, 6(1): 204-210.
%
%\bibitem{Mainardi1996}
%F. Mainardi. Fractional relaxation-oscillation and fractional
%diffusion-wave phenomena.
%\newblock \emph {Chaos, Solitons} \& \emph {Fractals}
%\newblock 1996, 7(9): 1461-1477.
%
%\bibitem{Podlubny1999}
%I. Podlubny.  \emph{Fractional differential equation}. San Diego,
%Academic press, 1999. 50-78.
%
%\bibitem{Podlubny2009a}
%I. Podlubny. Mittag-Leffler function.
%\newblock \emph {http://www.mathworks.de/matlabcentral /fileexchange/8738-mittag-leffler-function}
%\newblock 2009.
%
%\bibitem{Chen2008}
%Y. Q. Chen. Generalized Mittag-Leffler function.
%\newblock \emph {http://www.mathworks.de/matlabcentral/fileexchange/20849-generalized-mittag-leffler-function}
%\newblock 2008.
%
%\bibitem{Su2009}
%L. J. Su, W. Q. Wang, Z. X. Yang. Finite difference approximations
%for the fractional advection-diffusion equation.
%\newblock \emph {Phys. Lett. A}
%\newblock 2009, 373: 4405-4408.
%
%\bibitem{Smith2004}
%I. M. Smith, D. V. Griffiths.  \emph{Programming the finite element
%method (4th  edition) }. New York, John Wiley \& Sons Ltd, 2004.
%
%\bibitem{Meerschaert2004}
%M. M. Meerschaert, C. Tadjeran. Finite difference approximations for
%fractional advection-dispersion flow equations.
%\newblock \emph {Journal of
%Computational and Applied Mathematics}
%\newblock 2004, 172: 65-77.
%
%\bibitem{Reimus2003}
%P. Reimus, G. Pohll, T. Mihevc, J. Chapman, M. Haga, B. Lyles,
%S.Kosinski, R.Niswonger, P.Sanders, Testing and parameterizing a
%conceptual model for solute transport in a fractured granite using
%multiple tracers in a forced-gradient test.
%\newblock \emph {Water Resour. Res.}
%\newblock 2003, 39: 1356-1370.
%
%\bibitem{Pohll1999}
%G. Pohll, A. E. Hassan, J.B. Chapman, C. Papelis and R. Andricevic,
%Modeling ground water flow and radioactive transport in a fractured
%aquifer.
%\newblock \emph {Ground Water}
%\newblock 1999, 37(5): 770-784.
%
%\bibitem{Reimus2003}
%P. W. Reimus, M. J. Haga, A. I. Adams, T. J. Callahan, H. J. Turin,
%D. A. Coun, Testing and parameterizing a conceptual solute transport
%model in saturated fractured tuff using sorbing and nonsorbing
%tracers in cross-hole tracer tests.
%\newblock \emph {J. Contam. Hydrol.}
%\newblock 2003, 62-63: 613-636.
%
%\bibitem{Sze2008}
%K. Y. Sze, Y. K. Cheung. A hybrid-Trefftz finite element model for
%Helmholtz problem.
%\newblock \emph {Commun. Numer. Meth. Engng.}
%\newblock 2008, 24:2047-2060.

\end{thebibliography}
\end{document}